\documentclass[runningheads]{llncs}

\usepackage[T1]{fontenc}
\usepackage{graphicx}
\usepackage{amsmath}
\usepackage{amssymb}
\usepackage{booktabs}
\usepackage{multirow}
\usepackage{url}
\usepackage{hyperref}
\usepackage[utf8]{inputenc}

\begin{document}

\title{PriEval-Protect: A Unified Framework for Privacy Evaluation and Protection in Healthcare Systems}

\titlerunning{PriEval-Protect: Privacy Framework for Healthcare}

\author{Ilef Chebil\inst{1}\orcidID{0009-0002-3637-0567} \and
Asma Elhadj\inst{1}\orcidID{0009-0000-0388-6570} \and
Souheib Yousfi\inst{2}\orcidID{0000-0003-0810-0869} \and
Aroua Hedhili\inst{3}\orcidID{0000-0002-6918-0797} \and
Layth Sliman\inst{2}\orcidID{0000-0003-3369-7302}}

\authorrunning{I. Chebil et al.}

\institute{National Institute of Applied Science and Technology INSAT, Tunis, Tunisia\\
\email{ilef.chebil@insat.ucar.tn, asma.elhadj@insat.ucar.tn}
\and
Efrei Research Lab, Efrei, Paris-Panthéon-Assas University, Paris, France\\
\email{souheib.yousfi@efrei.fr, layth.sliman@efrei.fr}
\and
LARIA Research Lab, National School Of Computer Science ENSI, Tunis, Tunisia\\
\email{aroua.hedhili@ensi-uma.tn}}

\maketitle

\begin{abstract}
Safeguarding patient privacy while enabling meaningful healthcare data use remains critical under GDPR and HIPAA. Existing compliance methods are manual, error-prone, and separate policy audits from data-level assessments. This paper presents PriEval-Protect, a two-phase framework for unified privacy risk evaluation and mitigation. The evaluation phase combines regulatory compliance scoring using a fine-tuned legal LLM with RAG, and technical analysis via encryption type, data architecture, and metrics including similarity, uncertainty, adversary success, and information gain/loss. A composite risk score uses weighted aggregation via Analytic Hierarchy Process. The protection phase recommends countermeasures including federated learning and differential privacy based on assessed risk. Results on hospital documents and datasets demonstrate regulation-aligned, explainable assessments, bridging legal conformance and data-level risk analysis.

\end{abstract}

\section{Introduction}
\label{sec:introduction}

Healthcare systems increasingly rely on data-driven infrastructure for clinical decisions and research, yet must operate within strict regulatory boundaries like GDPR\footnote{\url{https://gdpr-info.eu/}} and HIPAA\footnote{\url{https://www.hhs.gov/hipaa/for-professionals/index.html}}. Current privacy evaluation remains fragmented and under-automated, with tools emphasizing either manual policy audits or isolated technical metrics without meaningful integration.

Regulatory frameworks like Digital Health Assessment Framework\footnote{\url{https://orchahealth.com/our-products/assessment-frameworks/digital-health-assessment-framework-dhaf/}} and HITRUST CSF\footnote{\url{https://hitrustalliance.net/hitrust-framework/}} offer structured evaluations but lack technical integration. Technical methods like Differential Privacy (DP)---adopted by Apple\footnote{\url{https://machinelearning.apple.com/research/scenes-differential-privacy}}---provide mathematical guarantees, while Federated Learning (FL) via frameworks like PySyft\footnote{\url{https://github.com/OpenMined/PySyft}}, Flower\footnote{\url{https://flower.ai/}}, and IBM FL\footnote{\url{https://github.com/IBM/federated-learning-lib}} enables decentralized training. However, these solutions fail to unify policy compliance with technical mitigation.

\paragraph{Our Contributions.}
\begin{itemize}
    \item \textbf{Unified Two-Phase Framework:} Combines (i) evaluation via regulatory scoring (fine-tuned legal LLM with RAG) and technical risk assessment (encryption, architecture, similarity, uncertainty, adversary success, information gain/loss), and (ii) protection recommending context-specific countermeasures (masking, Federated Learning FL, Differential Privacy DP) based on risk level.
    
    \item \textbf{Theoretical Integration:} Comprehensive privacy taxonomy integrating regulatory, architectural, and technical dimensions with Multi-Criteria Decision Making (MCDM) based risk aggregation using Analytic hierarchy process (AHP)\cite{taherdoost2023mcdm,saaty2008}.

    \item \textbf{Empirical Validation:} Evaluated on hospital policy documents and medical datasets, demonstrating regulation-aligned compliance scores and tailored protection strategies.
\end{itemize}

\section{Related Work}

Prior research divides into: (i) regulatory compliance, (ii) privacy evaluation (attribute classification and metrics), and (iii) protection techniques.

Regulatory compliance efforts like the Benjumea Privacy Scale \cite{benjumea2022} provide structured GDPR scoring via 14 weighted items validated by experts, but remain manual and error-prone. Privacy evaluation methods from Wagner and Eckhoff \cite{wagner2018} categorize 80+ metrics (uncertainty, information gain/loss, adversary success), yet lack healthcare-specific calibration. Tools like Pycanon \cite{pycanon2022} implement limited metrics requiring extensive preprocessing.

Attribute classification is essential for privacy metrics. Simões \cite{simoes2024} explored autoencoders, custom Deep Neural Networks (DNNs), and LLMs for Quasi-Identifiers/Sensitive Attributes (QID/SA) detection, showing promise but suffering from computational cost and interpretability issues. Two-step rule-based algorithms \cite{ruleclas2021} offer superior performance, transparency, and healthcare adaptability.

Protection techniques like FL \cite{fedlearn2023} and DP \cite{dwork2014} provide strong guarantees but lack context-aware application. Table~\ref{tab:relatedwork} summarizes existing approaches, highlighting gaps in integration between compliance, metrics, and adaptive protection that PriEval-Protect addresses.

\begin{table}
\centering
\caption{Comparative Summary of Existing Approaches}
\label{tab:relatedwork}
\small
\begin{tabular}{@{}p{2cm}p{2.2cm}p{2.5cm}p{2cm}@{}}
\toprule
\textbf{Criterion} & \textbf{Benjumea \cite{benjumea2022}} & \textbf{Wagner \cite{wagner2018}} & \textbf{Two-step \cite{ruleclas2021}} \\
\midrule
Domain & Compliance & Tech Metrics & Attribute Class. \\
Strengths & Weighted scoring & Comprehensive & Transparent \\
Limitations & Manual & No domain focus & Threshold tuning \\
\bottomrule
\end{tabular}
\end{table}

\begin{table}
\centering
\caption{Comparative Summary (Continued)}
\small
\begin{tabular}{@{}p{2.5cm}p{2.5cm}p{2.5cm}@{}}
\toprule
\textbf{LLM-based \cite{simoes2024}} & \textbf{FL, DP \cite{fedlearn2023,dwork2014}} \\
\midrule
Attribute Classification & Protection Techniques \\
Context-aware & Strong guarantees \\
Prompt-sensitive & Not adaptive \\
\bottomrule
\end{tabular}
\end{table}

\section{Background and Preliminaries}

\paragraph{Legal Frameworks.}
GDPR establishes EU data protection via principles like lawfulness, transparency (Art.5-6,13), purpose limitation, minimization, security (Art.32), impact assessments (Art.35), and breach notification (Art.33-34) \cite{amini2023ai}. HIPAA governs U.S. PHI privacy through technical, administrative, and physical safeguards \cite{smith2024hipaa}.

\paragraph{Data Classifications.}
Healthcare datasets contain: \textbf{Quasi-Identifiers (QIDs)}---indirect identifiers (age, ZIP); \textbf{Sensitive Attributes (SAs)}---directly sensitive fields (disease, income); \textbf{Non-Sensitive Attributes (NSAs)}---negligible risk attributes \cite{zhu2023multi,itm2021}.

\paragraph{Privacy Metrics.}
Quantify leakage risk \cite{wagner2018}: \textbf{Data Similarity (S)}---k-anonymity, l-diversity; \textbf{Uncertainty (U)}---entropy of sensitive attributes; \textbf{Adversary Success ($P_{\mathrm{adv}}$)}---re-identification probability; \textbf{Information Gain/Loss (IG/IL)}---mutual information or KL divergence changes.

\paragraph{AI Components.}
\textbf{LLMs} are transformer-based models for language understanding and generation \cite{zhao2023survey}, supporting compliance verification \cite{hassani2024compliance}. \textbf{RAG} merges LLMs with external retrieval for grounded responses \cite{lewis2020retrieval}. \textbf{Vector Databases} store embeddings for semantic search \cite{le2023vectordb}. \textbf{Fine-Tuning} adapts pre-trained LLMs via domain pretraining, supervised tuning, or parameter-efficient methods like LoRA with quantization \cite{anisuzzaman2025,ohri2024,wu2021,qlora2023}.

\paragraph{Privacy-Preserving Techniques.}
\textbf{Differential Privacy (DP)} quantifies privacy via $\Pr[\mathcal{M}(D_1) \in S] \leq e^\epsilon \cdot \Pr[\mathcal{M}(D_2) \in S] + \delta$ for neighboring datasets $D_1, D_2$, injecting calibrated noise \cite{dwork2014,ponomareva2023}. \textbf{Federated Learning (FL)} trains models across institutions without raw data exchange \cite{fedlearn2023,xu2021federated,antunes2022federated}.

\paragraph{MCDM.}
Multi-Criteria Decision Making provides structured frameworks for evaluating conflicting criteria. AHP translates qualitative judgments into quantitative weights via pairwise comparisons \cite{taherdoost2023mcdm,saaty2008}.

\section{Method}

\subsection{Framework Overview}

\begin{figure}
    \centering
    \includegraphics[width=\textwidth]{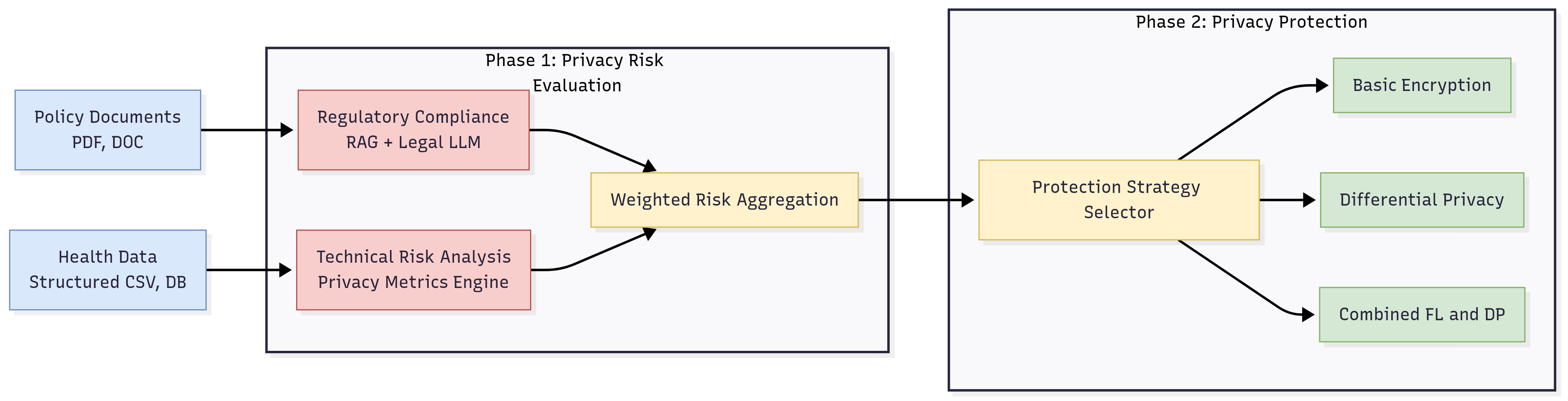}
    \caption{PrivEval-Protect Framework Overview.}
    \label{fig:pipeline}
\end{figure}

PriEval-Protect integrates regulatory and technical privacy evaluation in two phases. Phase 1 computes composite risk scores from policies and data; Phase 2 recommends tailored countermeasures. Figure~\ref{fig:pipeline} shows the pipeline: policy documents and health data are evaluated through legal LLMs and technical metrics, producing scores that guide protection strategies (DP, FL, or combined).

\subsection{Phase 1: Privacy Risk Evaluation}

\begin{figure}
    \centering
    \includegraphics[width=\textwidth]{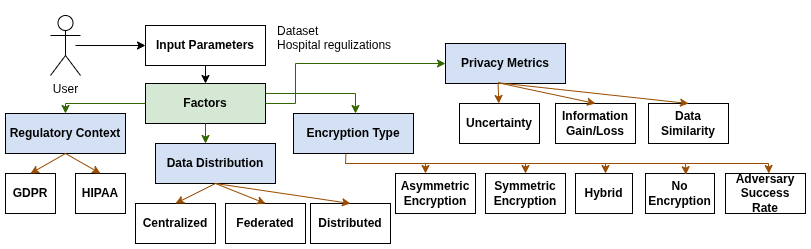}
    \caption{Privacy Risk Evaluation Taxonomy and Workflow.}
    \label{fig:global_framework}
\end{figure}

Phase 1 evaluates privacy via structured taxonomy covering regulatory, architectural, and technical factors (Figure~\ref{fig:global_framework}). Users upload datasets and policy documents, configuring encryption type (AES-256, RSA, hybrid) and data architecture (central, federated, distributed). The system computes privacy metrics and regulatory compliance scores, fused via weighted formula.

\subsubsection{Privacy Metrics Engine}

Custom Python implementations compute formal metrics from literature. After attribute classification via two-step rule-based algorithms \cite{ruleclas2021}, the module calculates: (1) Adversary Success Rate and Delta Presence, (2) k-Anonymity, alpha-k-Anonymity, l-Diversity, t-Closeness, (3) Information Gain and Conditional Entropy, (4) Shannon and normalized entropy. Multiprocessing optimizes large-scale processing.

\subsubsection{Regulatory Compliance Assessment}

Automated GDPR/HIPAA alignment combines fine-tuned LLM with RAG pipeline using Qdrant for semantic retrieval.

\paragraph{Dataset and Training.}
We curated: 9,000+ labeled GDPR policy segments, violation reports, and article corpus with Q\&A expansions. Preprocessing included normalization, article extraction, harmonization, and stratified oversampling (27:1 imbalance). Final corpus: 11,447 instances, 54\% compliant, 200+ indexed articles.

Two-stage fine-tuning: (1) Domain-Adaptive Pretraining on GDPR corpus using Saul-7B-Instruct, (2) Supervised Fine-Tuning on labeled examples with LoRA (6.8M parameters), 4-bit quantization, and gradient checkpointing.

\begin{figure}
    \centering
    \includegraphics[width=0.9\textwidth]{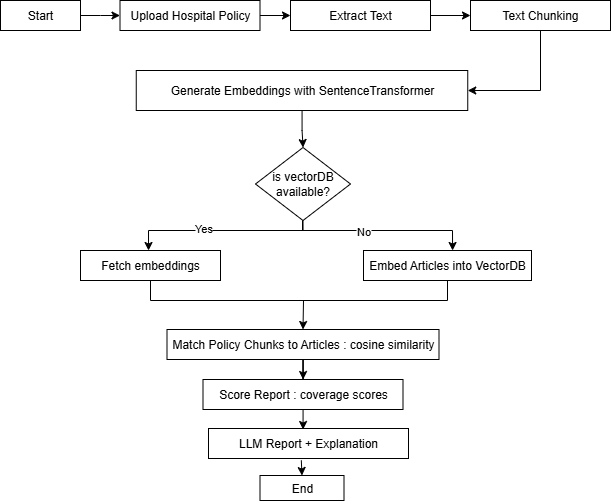}
    \caption{RAG-based compliance assessment pipeline.}
    \label{fig:rag_pipeline}
\end{figure}

\paragraph{RAG Pipeline and Scoring.}
Structure-aware chunking preserves legal context. Chunks are embedded (BAAI/bge-base-en-v1.5, L2-normalized) and stored in Qdrant. For each policy chunk, top-k GDPR articles are retrieved via cosine similarity. Compliance score:
\begin{equation}
    \text{Score} = \sum_{i=1}^{k}  s_i \cdot (1 - P_i)
\end{equation}
where $s_i$ is similarity and $P_i$ is penalty. Retrieved articles prompt the fine-tuned LLM, producing compliant/non-compliant labels with rationales and references (Figure~\ref{fig:rag_pipeline}).

\subsubsection{Score Aggregation}

Final privacy risk via weighted aggregation:
\begin{equation}
\text{Final Score} = \alpha \cdot C_{\text{compliance}} + \beta \cdot M_{\text{technical}} + \gamma \cdot E_{\text{encryption}} + \delta \cdot A_{\text{architecture}}
\end{equation}
Weights determined by AHP where $\alpha + \beta + \gamma + \delta = 1$. Risk categories: Low (80--100\%), Moderate (60--79\%), High (0--59\%).

\subsection{Phase 2: Privacy Protection}

Risk-based protection mapping (Table~\ref{tab:protection-strategies}): Low risk applies data masking; Moderate uses DP with noise injection; High activates FL+DP for collaborative training with encrypted updates.

\begin{table}
\centering
\caption{Privacy Protection Strategies}
\label{tab:protection-strategies}
\begin{tabular}{@{}llp{5.5cm}@{}}
\toprule
\textbf{Risk} & \textbf{Technique} & \textbf{Description} \\
\midrule
Low & Data Masking & Redacts sensitive fields \\
Moderate & Differential Privacy & Adds calibrated noise \\
High & FL + DP & Decentralized training with noise \\
\bottomrule
\end{tabular}
\end{table}

\section{Experimental Setup}

To empirically validate the PriEval-Protect framework, we conducted a multi-faceted evaluation combining real institutional data and expert assessments from healthcare organizations. The setup examines three key dimensions: (i) regulatory compliance scoring, (ii) technical privacy metrics computation, and (iii) unified privacy risk classification.

We utilized two categories of authentic institutional materials:

\begin{itemize}
  \item \textbf{Healthcare Policy Documents:} Internal privacy and data-sharing policies from multiple clinical institutions, used as input for the GDPR compliance assessment module. These documents represent diverse policy styles and governance maturity levels.
  \item \textbf{Clinical Data Samples:} Structured, anonymized patient datasets comprising demographic, clinical, and administrative fields. These were processed through our privacy metrics engine to quantify re-identification and inference risks.
\end{itemize}

For compliance scoring, we used the Benjumea Privacy Scale \cite{benjumea2022} as the baseline reference, aligning our LLM-based scoring model with established GDPR compliance indicators. For technical privacy metrics, the framework's computation engine was benchmarked against Pycanon \cite{pycanon2022} and ARX \cite{arx}, two widely recognized tools for evaluating anonymization and re-identification risks in data publishing. For global privacy risk classification, multiple privacy and data protection experts---including compliance officers and healthcare information specialists---were asked to manually assign risk levels (Low, Moderate, High) to various dataset scenarios. These labels were used as ground truth to validate the interpretability and accuracy of the aggregated PriEval-Protect risk scores.

\begin{table}
\centering
\caption{Evaluation criteria across PriEval-Protect components}
\label{tab:eval_metrics}
\begin{tabular}{@{}lp{7cm}@{}}
\toprule
\textbf{Component} & \textbf{Evaluation Metrics} \\
\midrule
Compliance Scoring & Mean Absolute Error (MAE), Pearson correlation \\
Privacy Metrics & Absolute deviation from PyCanon/ARX \\
Global Risk Score & Weighted agreement score \\
\bottomrule
\end{tabular}
\end{table}

\section{Results and Discussion}

To evaluate the effectiveness of PriEval-Protect, we performed a cross-component assessment covering compliance, technical, and integrated privacy risk evaluation. The quantitative and qualitative findings are summarized in Table~\ref{tab:evaluation-summary}.

\subsection{GDPR Compliance Assessment}

The compliance scoring module achieved a Mean Absolute Error (MAE) of 1.32 compared to expert-derived Benjumea Privacy Scale judgments, and a Pearson correlation of 0.78, indicating a strong positive linear relationship between predicted and expert scores. These results confirm that the LLM-based analysis reliably captures regulatory intent and contextual interpretation across diverse policy documents, validating its utility for semi-automated GDPR alignment.

\subsection{Technical Privacy Metric Benchmark}

The lightweight privacy metric engine demonstrated high consistency with formal tools. Deviations from Pycanon and ARX outputs were 0.6 and 0.4 respectively, indicating accurate computation of privacy measures such as k-anonymity, l-diversity, and t-closeness. This result highlights the framework's potential for integration into real-time data governance pipelines without compromising analytical rigor.

\subsection{Global Privacy Risk Validation}

Comparing system-generated global risk classes to expert annotations yielded an agreement rate of 82.6\%, with a recall of 81.2\% in high-risk cases---demonstrating that PriEval-Protect effectively identifies sensitive data configurations requiring stronger safeguards. These results suggest that the unified scoring model generalizes well beyond specific datasets or local contexts.

\subsection{Key Insights and Limitations}

The experimental outcomes validate PriEval-Protect as a robust framework capable of unifying legal and technical privacy evaluations. While results indicate strong agreement with human expertise, future studies should extend this validation to unstructured modalities (e.g., clinical narratives and imaging data) and broader institutional datasets. The framework's modular design makes it adaptable to varying compliance regimes, supporting its scalability across international healthcare ecosystems.

\begin{table}
\centering
\caption{Summary of Evaluation Results}
\label{tab:evaluation-summary}
\begin{tabular}{@{}llc@{}}
\toprule
\textbf{Component} & \textbf{Metric} & \textbf{Value} \\
\midrule
\multirow{2}{*}{Compliance} & MAE (vs. BPS) & 1.32 \\
 & Pearson Correlation & 0.78 \\
\midrule
\multirow{2}{*}{Privacy Metrics} & Deviation (PyCanon) & 0.6 \\
 & Deviation (ARX) & 0.4 \\
\midrule
\multirow{2}{*}{Global Score} & Class Agreement & 82.6\% \\
 & High-Risk Recall & 81.2\% \\
\bottomrule
\end{tabular}
\end{table}

\section{Conclusion}

PriEval-Protect advances privacy-aware AI in healthcare by integrating regulatory compliance, technical metrics, and data architecture analysis. It bridges legal, technical, and operational perspectives through LLM-driven scoring, weighted aggregation, and practical protection strategies (encryption, FL, DP). The framework enables automated, explainable privacy governance beyond manual audits. Future directions include extending to additional data types, enhancing automation, and real-world clinical deployment. PriEval-Protect demonstrates that scalable, automated privacy evaluation is achievable and necessary for responsible digital health innovation.

\begin{credits}
\subsubsection{\ackname}
This research was supported by INSAT, Efrei Research Lab, and LARIA Research Lab at ENSI.

\subsubsection{\discintname}
The authors have no competing interests to declare.
\end{credits}

\bibliography{references}

@article{taherdoost2023mcdm,
  title={Multi-Criteria Decision Making (MCDM) Methods and Concepts},
  author={Taherdoost, Hamed and Madanchian, Mitra},
  journal={Encyclopedia},
  volume={3},
  number={1},
  pages={Article 1},
  year={2023}
}

@article{saaty2008,
  title={Decision making with the analytic hierarchy process},
  author={Saaty, Thomas L},
  journal={International Journal of Services Sciences},
  volume={1},
  number={1},
  pages={83--98},
  year={2008}
}

@article{benjumea2022,
  title={Privacy Scale for mobile health applications},
  author={Benjumea, J and others},
  journal={JMIR mHealth and uHealth},
  volume={10},
  number={7},
  pages={e37591},
  year={2022}
}

@article{wagner2018,
  title={Technical privacy metrics: A systematic survey},
  author={Wagner, Isabel and Eckhoff, David},
  journal={ACM Computing Surveys},
  volume={51},
  number={3},
  pages={1--38},
  year={2018}
}

@article{pycanon2022,
  title={A Python library to check the level of anonymity of a dataset},
  author={S{\'a}inz-Pardo D{\'i}az, Judith and L{\'o}pez Garc{\'i}a, {\'A}lvaro},
  journal={Scientific Data},
  volume={9},
  pages={785},
  year={2022}
}

@phdthesis{simoes2024,
  title={Automated Attribute Classification for Privacy Metrics},
  author={Sim{\~o}es, Ant{\'o}nio},
  school={University of Coimbra},
  year={2024}
}

@article{ruleclas2021,
  title={Two-step rule-based classification algorithm for privacy-preserving data mining},
  journal={Security and Communication Networks},
  pages={7154705},
  year={2021}
}

@article{fedlearn2023,
  title={Federated Learning applications in healthcare: A systematic review},
  journal={International Journal of Environmental Research and Public Health},
  volume={20},
  number={15},
  pages={6539},
  year={2023}
}

@article{dwork2014,
  title={The algorithmic foundations of differential privacy},
  author={Dwork, Cynthia and Roth, Aaron},
  journal={Foundations and Trends in Theoretical Computer Science},
  volume={9},
  number={3-4},
  pages={211--407},
  year={2014}
}

@article{amini2023ai,
  title={AI Ethics and Challenges in Healthcare under GDPR},
  author={Amini, Mohammad Mahdi and others},
  journal={Machine Learning and Knowledge Extraction},
  volume={5},
  number={3},
  pages={1023--1035},
  year={2023}
}

@article{smith2024hipaa,
  title={Cybersecurity in Healthcare and HIPAA Compliance},
  author={Smith, David A and Abbasi, Noman},
  journal={Journal of Knowledge, Learning and Science and Technology},
  volume={3},
  number={3},
  pages={278--287},
  year={2024}
}

@article{zhu2023multi,
  title={Multiple sensitive attributes publishing with guaranteed utility},
  author={Zhu, Haoran and others},
  journal={CAAI Transactions on Intelligence Technology},
  volume={8},
  number={2},
  pages={288--296},
  year={2023}
}

@inproceedings{itm2021,
  title={Privacy-preserving healthcare informatics: a review},
  booktitle={ICMSA 2021},
  series={ITM Web of Conferences},
  year={2021}
}

@article{zhao2023survey,
  title={A Survey of Large Language Models},
  author={Zhao, Wayne Xin and others},
  journal={arXiv preprint arXiv:2303.18223},
  year={2023}
}

@inproceedings{hassani2024compliance,
  title={Enhancing Legal Compliance and Regulation Analysis with LLMs},
  author={Hassani, Sepehr},
  booktitle={RE 2024},
  pages={507--511},
  address={Reykjavik},
  year={2024}
}

@inproceedings{lewis2020retrieval,
  title={Retrieval-augmented generation for knowledge-intensive NLP},
  author={Lewis, Patrick and others},
  booktitle={Advances in Neural Information Processing Systems},
  volume={33},
  pages={9459--9474},
  year={2020}
}

@article{le2023vectordb,
  title={A Survey on Vector Databases},
  author={Le, Minh and others},
  journal={arXiv preprint arXiv:2310.11703},
  year={2023}
}

@article{anisuzzaman2025,
  title={Fine-Tuning Large Language Models for Specialized Use Cases},
  author={Anisuzzaman, D M and others},
  journal={Mayo Clinic Proceedings: Digital Health},
  volume={3},
  number={1},
  pages={100184},
  year={2025}
}

@article{ohri2024,
  title={Supervised fine-tuned approach for automated detection of diabetic retinopathy},
  author={Ohri, Kanika and Kumar, Manoj},
  journal={Multimedia Tools and Applications},
  volume={83},
  pages={14259--14280},
  year={2024}
}

@inproceedings{qlora2023,
  title={QLoRA: Efficient Finetuning of Quantized LLMs},
  author={Dettmers, Tim and others},
  booktitle={Advances in Neural Information Processing Systems},
  volume={36},
  pages={10088--10115},
  year={2023}
}

@article{xu2021federated,
  title={Federated Learning for Healthcare Informatics},
  author={Xu, Jie and others},
  journal={Journal of Healthcare Informatics Research},
  volume={5},
  pages={1--19},
  year={2021}
}

@article{antunes2022federated,
  title={Federated Learning for Healthcare: Systematic Review and Architecture Proposal},
  author={Antunes, Rodrigo S and others},
  journal={ACM Transactions on Intelligent Systems and Technology},
  volume={13},
  number={4},
  pages={54:1--54:23},
  year={2022}
}

@article{wu2021,
  title={Domain-Adaptive Pretraining Methods for Dialogue Understanding},
  author={Wu, Hao and others},
  journal={arXiv preprint arXiv:2105.13665},
  year={2021}
}

@article{ponomareva2023,
  title={How to DP-fy ML: A practical guide to machine learning with differential privacy},
  author={Ponomareva, Natalia and others},
  journal={Journal of Artificial Intelligence Research},
  volume={77},
  pages={1113--1201},
  year={2023}
}

@incollection{arx,
  title={Putting Statistical Disclosure Control into Practice: The ARX Data Anonymization Tool},
  author={Christen, Peter and others},
  booktitle={Medical Data Privacy Handbook},
  pages={111--148},
  year={2015}
}

\end{document}